\documentclass[pre,twocolumn,groupedaddress,showpacs,showkeys]{revtex4}

	\usepackage{amssymb}
	\usepackage{amsmath}
	\usepackage{graphicx}
	\usepackage{subfigure}
	\usepackage{dsfont}
	\usepackage{mathrsfs}
	\usepackage{natbib}       
	\usepackage[dvips]{color} 
\newcommand{\fix}[1]			
             {{\color{magenta} #1 }}

\newcommand{\beq}{\begin{equation}}
\newcommand{\eeq}{\end{equation}}
\newcommand{\ee}[1] {\label{#1} \end{equation}}
\newcommand{\bea}{\begin{eqnarray}}

\newcommand{\eea}{\end{eqnarray}}

\begin{document}

\title{
Scaling behavior of domain walls at the $T=0$ ferromagnet to spin-glass transition
}
\author{
O.~Melchert and A.~K.~Hartmann
}
\affiliation{
Institut f\"ur Physik, Universit\"at Oldenburg, 26111 Oldenburg, Germany
}
\date{\today} 

\begin{abstract}
We study domain-wall excitations in two-dimensional random-bond Ising spin
systems on a square lattice with side length $L$, 
subject to two different continuous disorder distributions.
In both cases an adjustable parameter allows to tune the disorder
so as to yield a transition from a spin-glass ordered ground state 
to a ferromagnetic groundstate.
We formulate an auxiliary graph-theoretical problem in which domain 
walls are given by undirected shortest paths with possibly negative
distances.
Due to the details of the mapping, standard shortest-path algorithms 
cannot be applied. 
To solve such shortest-path problems we have to apply
minimum-weight perfect-matching algorithms. 
We first locate the critical values of the disorder parameters, 
where the ferromagnet to spin-glass transition occurs for the two types of the
disorder. 
For certain values of the disorder parameters close to the respective critical 
point, we investigate 
the system size dependence of the width of the the average 
domain-wall energy ($\sim\!L^{\theta}$)
and the average domain-wall length ($\sim\!L^{d_{\rm{f}}}$).
Performing a finite-size scaling analysis for systems with a side 
length up to $L\!=\!512$,
we find that both exponents remain constant in the spin-glass phase, 
i.e.\ $\theta\!\approx\!-0.28$ and $d_{\rm f}\!\approx\!1.275$. 
This is consistent 
with conformal field theory, where it seems to be possible to relate the 
exponents from the analysis of Stochastic Loewner evolutions (SLEs)
via $d_{\rm f}-1\! =\!3/[4(3+\theta)]$.
Finally, we characterize the transition in terms of ferromagnetic clusters of 
spins that form, as one proceeds from spin-glass ordered to ferromagnetic
ground states.
\end{abstract} 


\maketitle

\section{Introduction}
Ising spin glasses (ISGs) are among the most-basic models of
disordered systems that allow for the study of phase transitions in the
presence of quenched disorder. ISGs are elaborately  studied in statistical
physics \cite{binder1986,fischer1991,mezard1987,young1998}  and despite
several decades of active  research they attract a constant interest,
challenging with still not well understood traits and unresolved questions.
In the scope of this paper we investigate ground state (GS) spin configurations
and minimum-energy domain-wall (MEDW) excitations in a $2d$ random-bond ISG.  
In brief, MEDWs are topological excitations that are induced by a change of the  
boundary conditions (BCs) from periodic to antiperiodic along one boundary of the system.  
In particular we are interested in the scaling properties of MEDWs close to the critical
point at which the $T\!=\!0$ spin glass (SG) to ferromagnet (FM) transition
occurs.  From a phenomenological point of view, the physics of the SG ordered
phase of ISGs with  short ranged interactions, like the $2d$ model considered
here, can be described in terms of the droplet scaling picture
\cite{mcmillan1984,fisher1986,fisher1988}.  Therein, the low-temperature
behavior is dominated by droplet excitations, i.e.\ clusters of spins that are
flipped relative to the GS spin configuration.  Within the
droplet picture, excitations like MEDWs posses an excitation energy $\Delta E$
that scales with  system size $L$ as $\Delta E\sim L^{\theta}$, where $\theta$
is referred to as stiffness  exponent. The value of $\theta$ is assumed to be
universal for all types of excitations and  constant within the whole SG
phase.  Moreover, in a statistical sense, they are self-similar fractals
characterized by  a fractal dimension 
$d_{\rm f}$ that is defined by the
scaling of the average MEDW length as $\langle l \rangle\!\sim\!L^{d_{\rm
f}}$.  The advantage of working at zero temperature is  that the GS problem
for the particular setup studied here can be solved by means of exact
combinatorial-optimization algorithms
\cite{SG2dReview2007,opt-phys2001,bieche1980,barahona1982,pardella2008} whose
running time increases only polynomially with the system size. Hence, very
large systems  can be treated exactly, giving very precise and reliable
estimates for the observables.  For $2d$ lattices, where the interaction
strengths (bonds) between adjacent spins  are drawn from a Gaussian
distribution with zero mean and unit width, domain wall (DW) 
calculations  using
such algorithms  resulted in the estimates $\theta\!=\!-0.287(4)$
\cite{hartmann2001,hartmann2002} and $d_{\rm f}\!=\!1.274(2)$
\cite{melchert2007}.  The negative value of the stiffness exponent indicates
that the excitation energy required to  introduce a MEDW gets negligibly small
as $L\rightarrow \infty$ and thus,  thermal fluctuations prevent a spin-glass
ordering for any non-zero temperature.  The above value of the stiffness
exponent was later on confirmed for continuous disorder distributions
different from the Gaussian bond distribution \cite{amoruso2003}, for droplet
excitations  respecting a Gaussian distribution of the bonds
\cite{hartmann2002b,droplets2003,hartmann2004} 
and quite recently also for droplets within
the $\pm J$ model \cite{hartmann2008}.  Furthermore, recent studies suggested
that MEDWs respecting a Gaussian distribution of the bonds can be described by
stochastic Loewner evolutions (SLEs) \cite{amoruso2006,bernard2007}.  SLEs are
generated by a stochastic differential equation driven by a brownian motion.
They describe the continuum limit for various $2d$ random curves and their
geometric  properties relate to the statistics of several critical interfaces
\cite{cardy2005}.  Within conformal field theory  it further seems to be
possible to relate the DW fractal dimension to the stiffness exponent
by means of the relation $d_{\rm f}\!-\!1\!=\!3/[4(3\!+\!\theta)]$,
subsequently referred to as SLE scaling relation.  For the pure spin glass,
this is in agreement with the numerical estimates of $\theta$ and $d_{\rm f}$
stated above.

Here, we consider a random-bond Ising model that allows us to investigate the
SG to FM transition at zero temperature, by tuning the mean value of the
underlying  disorder distribution. 
In a previous work, the related $\pm J$ model was studied 
in $2d$ \cite{amoruso2003}. There exact matching algorithms
to find ground states (GSs) where applied. 
It was found that, in the limit of large system sizes, the SG to FM transition
occurs at a fraction $p_{\rm c}=0.103(1)$ of antiferromagnetic bonds ($-J$)
among ferromagnetic bonds ($+J$). Further, the critical exponents $\nu$ and
$\beta$ that describe the divergence of the correlation length and the order
parameter,  where found to be $\nu=1.55(1)$ and $\beta=0.09(1)$.
Due to the discreteness of the distribution, the DWs are not unique
and cannot be sampled in equilibrium for large systems. Hence the
fractal dimension has not been determined in a precise way so far.

To clarify whether the SLE scaling
relation above holds within the whole  spin-glass phase, we use two
different continuous distributions of the disorder, which allow us
to calculate the fractal dimension $d_{\rm f}$ with high
precision. For this purpose, we perform GS
calculations by means of exact combinatorial-optimization  algorithms and
study the scaling behavior of MEDWs close to the critical point where the SG
to FM transition occurs.  At first, we perform a finite-size scaling analysis
for systems of moderate sizes ($L\!\leq\!64$)  to locate the critical points at
which the transitions takes place.  Then we perform additional simulations for
large systems ($L\!\leq\!512$) close to  and directly at the critical points,
to get a grip on the scaling behavior of the MEDWs.  
Finally, we characterize the transition using a finite-size scaling analysis 
for the largest and second-largest ferromagnetic clusters of spins within the 
GS spin configurations. These clusters form as one proceeds from 
spin-glass ordered to ferromagnetic ground states.
To summarize our results:
we find that the SLE scaling relation holds in the SG phase up to a point very
close  to the respective critical points, but not right at the critical
points. Moreover, MEDWs in the SG phase scale
like self-similar  fractals, while MEDWs in the ferromagnetic phase display a
self-affine scaling behavior.  

The paper is organized as follows. In section
\ref{sec:model} we introduce the model and describe the algorithmic techniques
we have used in order to obtain MEDWs.  In section \ref{sec:results} we
present the results of our numerical simulations.  We conclude with a summary
in section \ref{sec:summary}.


\section{Model and Method}
\label{sec:model}
We performed GS calculations for
two-dimensional random-bond Ising spin systems with nearest-neighbor
interactions. The respective model consists of $N=L \times L$ spins
$\sigma=(\sigma_1,\ldots,\sigma_N)$ with $\sigma_i=\pm 1$, located on
the sites of a regular square lattice. The energy of a given spin
configuration is measured by the Edwards-Anderson Hamiltonian
\begin{eqnarray}
H(\sigma) = -\sum_{\langle i,j \rangle} J_{ij} ~\sigma_i \sigma_j,
\label{eq:EA_hamiltonian}
\end{eqnarray}
where the sum runs over all pairs of adjacent spins with periodic 
BCs in the $x$--direction and free BCs in the $y$--direction.  Therein, the
bonds $J_{ij}$ are quenched random variables drawn from a given disorder distribution.
Subsequently, we distinguish two types of the bond disorder:
%
\begin{figure}[t!]
\centerline{
\includegraphics[width=1.0\linewidth]{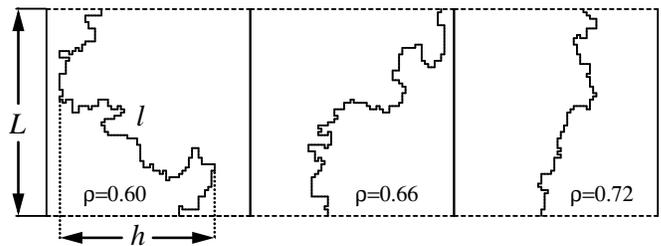}} 
\caption{
Domain wall samples for systems of side-length $L\!=\!64$ and different
values of the disorder parameter $\rho$ (Model I). The samples are taken
in the SG phase ($\rho\!=\!0.60$), right at the critical point ($\rho\!=\!0.66$) and
in the FM phase ($\rho\!=\!0.72$). Besides the system size $L$, the DW length $l$
and its roughness $h$ are illustrated.
\label{fig1}}
\end{figure}  
%

(1) Model I, where one realization of the disorder consists of a random
fraction  $\rho$ of ferromagnetic bonds and a fraction $(1-\rho)$ of bonds
that are drawn from a Gaussian distribution with zero mean and unit variance,
i.e.\
\begin{eqnarray}
P_{\rm I}(J)\!&=&\!(1\!-\!\rho)~\exp(-J^2/2)/\sqrt{2\pi}~+~\rho~\delta(J\!-\!1).
\end{eqnarray}
There exists a critical value $\rho_{\rm c}$ of the disorder parameter that
separates  a spin-glass phase ($\rho<\rho_{\rm c}$) from a ferromagnetic phase
($\rho>\rho_{\rm c}$).  As limiting cases we can identify the pure Ising SG at
$\rho=0$ and the ordinary  Ising ferromagnet at $\rho=1$.  A similar type of
disorder was used earlier for Monte Carlo simulations that where  carried out
to study the FM to SG transition in $3d$ and to numerically verify the absence
of an equilibrium  ``mixed'' ferromagnetic-SG phase for the respective model
\cite{krzakala2002}.  

(2) Model II, where the bond strengths are drawn 
from a Gaussian distribution with
mean $\mu_J$ and width $\sigma_{J}$, i.e.\
\begin{eqnarray}
P_{\rm II}(J)\!&=&\!\exp(-(J-\mu_J)^2/(2 \sigma_J^2))/(\sqrt{2\pi}\sigma_J).
\end{eqnarray}
As a function of the reduced 
variable $r=\sigma_J/\mu_J$ we expect to find a ferromagnetic phase 
(spin-glass phase) for $r\!<\!r_{\rm c}$ ($r\!>\!r_{\rm c}$).
An earlier DW renormalization-group study of small systems
\cite{mcmillan1984b} supported by transfer-matrix 
calculations reported, amongst other things, a zero temperature FM to SG 
transition at $r_{\rm c}=0.961(10)$ with $\nu=1.42(8)$. Further, for the pure SG
($\mu_J\!=\!0$), an extrapolation of the DW free energy to zero 
temperature resulted in a stiffness exponent $\theta=-0.281(5)$. 
The fractal properties of the DWs were not studied in this work.

In the above two models, the bonds are allowed to take either sign, where a value 
$J_{ij}\!>\!0$ signifies a ferromagnetic coupling that prefers a parallel alignment of 
the coupled spins, while a value $J_{ij}\!<\!0$ indicates an antiferromagnetic 
coupling in favor of antiparallel aligned spins.
The competing nature of these interactions gives rise to frustration. A plaquette, i.e.\ an
elementary square on the lattice, is said to be frustrated if it is
bordered by an odd number of antiferromagnetic bonds. In effect, frustration
rules out a GS in which all the bonds are satisfied. 

Here, our intention is to get a grip on the geometric properties of
minimum-energy DWs.  These are topological excitations that are
defined, for each realization of the bond disorder, relative to two spin
configurations: $\sigma_p$, a GS spin configuration with respect to periodic
BCs further characterized by the configurational energy $E_p$ and
$\sigma_{ap}$, a GS respecting antiperiodic BCs characterized by the energy
$E_{ap}$.  Antiperiodic BCs are realized by inverting the sign of all the
bonds along one column  in the $x$-direction.  Comparing the orientation of
the spins in the two GSs, one can distinguish two regions on the lattice: one
where the orientation of a spin is the same in both GSs and another, where the
orientation of a spin differs regarding the two GSs. Within these regions, the
bonds between  adjacent spins are either satisfied or broken in both GSs
likewise. Bonds that connect spins that belong to different regions on the
lattice are satisfied in exactly one of the two GSs.  The MEDW is the
interface in between the two regions and as such, it runs perpendicular to the
latter bonds.  It has the property that its excitation energy  $\delta
E=E_{ap}-E_p$ is minimal among all possible DWs that span the system in the
direction with the free BCs.  The basic observables related to a DW
are its over all length $l$, its roughness $h$ and its excitation energy
$\delta E$.  MEDWs for three different values of the disorder parameter $\rho$
introduced above (see Model I) are  illustrated in figure \ref{fig1}.

We now give a brief description of the algorithm that we used to determine the
MEDWs.  A more extensive description of the individual steps of the algorithm
can be found in \cite{melchert2007}.  For a given realization of the bond
disorder, we first determine a GS spin configuration  consistent with periodic
BCs in the $x$-direction. Besides the  magnetization $m_L=|\sum_i \sigma_i|/
L^{2}$ and the energy, this tells  which bonds are satisfied/broken in the GS
for that particular disorder sample.  For the $2d$ ISG on planar lattice
graphs, i.e.\ when there are periodic BCs in at most one direction, exact GS
spin configurations  can be found in polynomial time.   This is possible
through a mapping to an appropriate  minimum-weight perfect-matching problem
\cite{opt-phys2001,bieche1980,SG2dReview2007}, a combinatorial-optimization
problem known from computer science.  Here, we state only the general idea of
this method.  For this mapping, the spin system needs to be represented by its
frustrated plaquettes and paths connecting those pairwise, i.e.\ {\em
matching} them. In doing so, individual path segments are confined to run
perpendicular across bonds on the spin lattice.  Those bonds that are crossed
by path segments are not satisfied in  the corresponding spin configuration.
The {\em weight} of the matching is just the sum of the absolute values of all
bond strengths that relate to unsatisfied bonds.  Hence, finding a
minimum-weight perfect matching on the graph of frustrated plaquettes then
corresponds to finding a spin configuration with a minimal  configurational
energy, hence a GS.  The use of this approach permits the treatment of large
systems, easily up to $L=512$, on single processor systems.  This GS spin
configuration can further be used to set up a weighted dual of the  spin
lattice, whose weighted edges comprise all possible DW segments.  The weighted
dual is constructed as follows: set up a new graph $G\!=\!(V,E,\omega)$,
whose sites $i\in V$ relate to the elementary plaquettes on the spin
lattice. Its necessary to  introduce two {\em extra} sites that account for
the free BCs. Two sites are joined by an undirected edge $e\in E$, if the
corresponding plaquettes have a bond in common.  For the weight assignment on
the dual, consider a bond on the spin lattice  having a coupling strength
$J_{ij}$. If the bond is satisfied (broken) regarding the GS, the
corresponding dual edge $e$ gets a weight $\omega(e)=-2|J_{ij}|$
($\omega(e)=+2|J_{ij}|$).  The weighted dual now comprises all possible DW
segments, where  the weight of an edge is equal to the amount of energy that
it would contribute  to a DW. Every possible DW links both extra sites on the
dual, where the energy of a  DW is the sum of the weights along the according
lattice path.  So as to have minimum energy, it is beneficial for a DW to
include (avoid) edges with a  negative (positive) edgeweight.  Consequently, a
MEDW is a minimum-weight path on the dual that joins both extra sites.  The
dual $G$ is an undirected graph that allows for negative edge weights and so
as  to construct minimum-weight paths on $G$, it requires matching techniques
\cite{ahuja1993}.  Therefore we need to map the dual to an auxiliary graph
$G_{\rm A}$ and find a minimum-weight perfect matching on $G_{\rm A}$ which we
finally can relate to a minimum-weight path on $G$.  For each realization of
the disorder, this procedure yields an explicit representation of the
minimum-energy DW that we can easily probe for its geometric properties.  A
more detailed description of the algorithm can be found in
\cite{melchert2007}.  In the following we will use the procedure outlined
above to investigate MEDWs for  the random-bond Ising models introduced
earlier.

\section{Results}
\label{sec:results}
So as to characterize the scaling behavior of MEDWs for 
the two disorder distributions introduced above,
we first of all need to find the critical values $\rho_{\rm{c}}$ (Model I) 
and $r_{\rm{c}}$ (Model II) of the disorder parameters
at which the $T\!=\!0$ SG to FM transition takes place.
Reliable estimates for the location of the critical points can
already be obtained from comparatively small system sizes, here we 
use $L\!=\!24,32,48,64$. In general, one has to find a proper balance
of system size and sample numbers that affect finite-size effects and 
statistical error, respectively \cite{newman00}.
Subsequently we can probe the asymptotic scaling behavior of the MEDWs at fixed
values of the disorder parameters close to the critical points for
large system sizes up to $L\!=\!512$.

%
\begin{figure}[t!]
\centerline{
\includegraphics[width=1.0\linewidth]{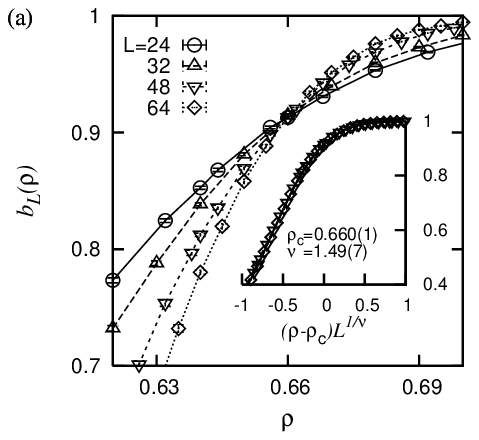}} 
\centerline{
\includegraphics[width=1.0\linewidth]{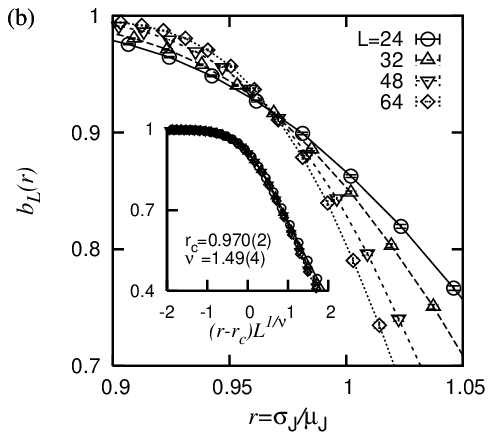}} 
\caption{Results of the FSS analysis for the Binder parameter 
$b_L$ associated with the magnetization, for different 
system sizes $L$. The main plot shows the unscaled data near 
the critical point, while the inset illustrates the data 
collapse obtained for 
(a) Model I:  $\rho_{\rm{c}}\!=\!0.660(1)$ and $\nu\!=\!1.49(7)$,
(b) Model II: $r_{\rm{c}}\!=\!0.970(2)$ and $\nu\!=\!1.49(4)$.
\label{fig2ab}}
\end{figure}  
%
\subsection{Finite-size scaling analysis to characterize the $T\!=\!0$ spin glass to 
ferromagnet transition}
\begin{figure}[t!]
\centerline{
\includegraphics[width=1.0\linewidth]{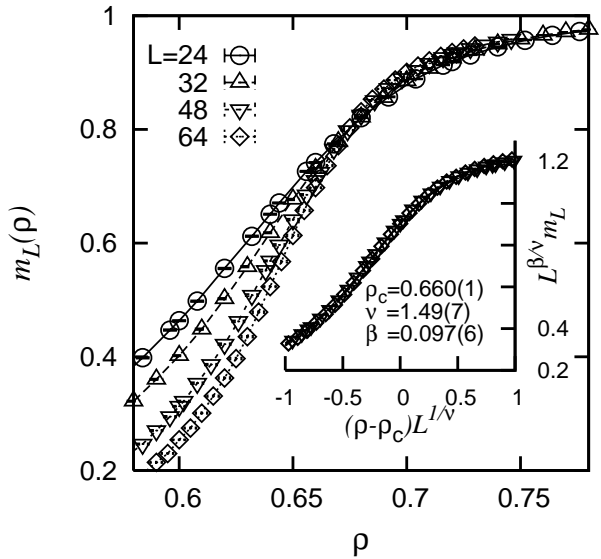}} 
\caption{Results of the FSS for the average magnetization 
$m_L(\rho)$ for different system sizes $L$ for Model I 
disorder. 
The main plot shows the unscaled data near the critical point,
while the inset illustrates the data collapse obtained 
for the parameters $\rho_{\rm{c}}\!=\!0.660(1)$, $\nu\!=\!1.49(7)$
and $\beta\!=\!0.097(6)$.
\label{fig3}}
\end{figure}

\begin{figure}[t!]
\centerline{
\includegraphics[width=1.0\linewidth]{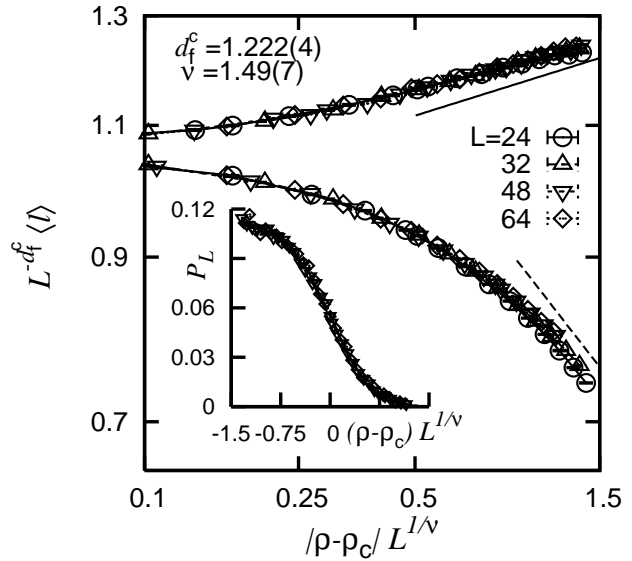}} 
\caption{Results of the FSS analysis for Model I disorder.
The main plot 
illustrates the FSS of the average MEDW length 
$\langle l \rangle$ for different  
system sizes $L$, where the best data collapse is obtained 
for the parameters $\rho_{\rm{c}}\!=\!0.660(1)$, $\nu\!=\!1.49(7)$
and $d_{\rm f}\!=\!1.222(4)$. The solid an dashed lines illustrate
the asymptotic scaling behavior of both branches as described in 
the text. The inset shows the 
scaling of the probability $P_L(\rho)$ that the roughness of
the MEDW is equal to the system size $L$.
\label{fig4}}
\end{figure}

First, we will discuss the results for the Model I disorder and afterwards report 
the results for the Model II disorder more briefly.
As pointed out above, at large values of $\rho$, there exists
an ordered ferromagnetic
phase, while for small values of $\rho$ a spin-glass phase exists. Therefore, 
a proper order parameter to characterize the respective SG-FM transition is the 
magnetization $m_L=|\sum_i \sigma_i|/ L^{2}$
 for a system of size $L$. 
In the following, we perform a finite-size scaling analysis (FSS) in order
to locate the critical point $\rho_{\rm c}$ and also estimate the
critical exponents that describe the scaling behavior of the magnetization at 
criticality.
The Binder parameter \cite{binder1981} associated with the magnetization reads 
\begin{eqnarray}
b_L\!=\!\frac{1}{2}\Big(3-\frac{\langle m_L^4\rangle}{\langle m_L^2\rangle^2 }\Big)
\end{eqnarray}
and is expected to scale as
$b_L(\rho)\!\sim\!f_1[ (\rho-\rho_{\rm c}) L^{1/\nu}]$,
where $f_1$ is a size-independent function and
 $\nu$ signifies the critical 
exponent that describes the divergence of the correlation length as the 
critical point is approached. Here, we simulated square systems of size $L=24,32,48,64$
at various values of the disorder parameter $\rho$. Observables are averaged over up to
$3\!\times\!10^4$ ($2\!\times\!10^4$) samples for the smallest (largest) systems and we
utilized the data collapse anticipated by the scaling assumption above to obtain 
$\rho_{\rm c}\!=\!0.660(1)$ and $\nu\!=\!1.49(7)$ with a quality 
$S\!=\!1.25$ of the data collapse \cite{houdayer2004}, see figure \ref{fig2ab}(a).
The value of the critical exponent $\nu$ agrees within errorbars with the 
value $\nu\!=\!1.42(8)$ obtained using a transfer-matrix approach \cite{mcmillan1984b}.
Note that both, the numerical values of $\rho_{\rm c}$ and $\nu$ further agree 
with those that characterize the negative-weight percolation of
loops and paths on $2d$ lattices \cite{melchert2008}, highlighting the close connection of the 
two optimization problems. 
The order parameter of the transition is expected to scale conform 
with the assumption
$m_L(\rho)\!\sim\!L^{-\beta/\nu} f_2[(\rho-\rho_{\rm c}) L^{1/\nu}],$
$f_2$ being a size-independent function,
where the magnetization exponent $\beta$ was obtained 
after fixing $\nu$ and $\rho_{\rm c}$ to the values stated above. 
The most satisfactory data collapse ($S\!=\!1.83$) was obtained using 
$\beta\!=\!0.097(6)$, see figure \ref{fig3}. 
In general, the above scaling relation holds best near the critical point and 
one can expect that there are corrections to scaling off criticality. 
As a remedy, we restricted the latter scaling analysis to the interval 
$[-0.5,+0.2]$, enclosing the critical point on the rescaled abscissa. 
Note that the values for the exponents found here agree with those found from 
GS calculation for the $\pm J$--model \cite{amoruso2004} within the errorbars.
Further, the exponents appear to be consistent with those that describe the paramagnet
to ferromagnet transition for the random bond Ising model, regarding finite temperatures $T<T^*$
below the temperature $T^*$ that characterizes the multicritical Nishimori point \cite{toldin2009}. 
In the respective study, the exponents $\nu\!=\!1.50(4)$ and $\beta\!=\!0.095(5)$ where
measured by means of monte carlo simulations for the random bond $\pm J$ Ising model 
on lattices with $L\!\leq\!64$ at fixed $T$, while varying the 
fraction of ferromagnetic bonds on the lattice.
As an alternative order parameter, we also studied the average path length $\langle l \rangle$
of the MEDWs, where we expect a scaling of the form
\begin{equation} 
\langle l \rangle\!\sim\!L^{d_{\rm f}^{\rm c}}f_3[(\rho-\rho_{\rm{c}})L^{1/\nu}]\,.
\label{eq:scaling:l}
\end{equation}
Therein, $d_{\rm f}^{\rm c}$ signifies the fractal dimension of the DWs at  the
critical point and $f_3$ is another size-independent function.  From
a finite-size scaling analysis restricted to the interval $[-0.75,+0.5]$ on
the rescaled abscissa, we obtained  $d_{\rm f}^{\rm c}\!=\!1.222(4)$ with a
quality $S\!=\!1.33$, see figure \ref{fig4} (Note that for a more clear
presentation, the argument along the abscissa in figure \ref{fig4} reads
$|\rho-\rho_{\rm{c}}|L^{1/\nu}$). For the somewhat larger interval  $[-1,+0.5]$ we
found $d_{\rm f}^{\rm c}\!=\!1.223(4)$ with $S\!=\!1.40$  in agreement  with
the above value.  Since we expect the average MEDW length at $\rho\!=\!0$ to
scale as  $\langle l \rangle\!\sim\! L^{d_{\rm f}}$ (here, $\rho\!=\!0$
corresponds to the  pure spin glass studied in \cite{melchert2007}), where
$d_{\rm f}\!=\!1.274(2)$, we can further estimate the asymptotic behavior
$f_3(x)\sim x^{\nu (d_{\rm f}-d_{\rm f}^{\rm c})}$  of the scaling function 
in Eq.\ (\ref{eq:scaling:l}) as
$x\!\rightarrow\! -\infty$. This can be seen from the top branch in figure
\ref{fig4}, where the function $f_3(x)\!\sim\! x^{0.08(1)}$ is shown as solid
line and agrees well with the data. Note that via 
Eq. (\ref{eq:scaling:l}) the DWs at $\rho_{\rm{c}}$ exhibit the fractal dimension
$d_{\rm f}^{\rm c}$, while for {\em all} values $\rho<\rho_{\rm{c}}$, the
fractal dimension is given by $d_{\rm f}$. Hence, the scaling
ansatz Eq. (\ref{eq:scaling:l}) is based on the assumption that
behavior in the SG phase is universal, which is tested below for much larger
systems explicitly.

 For the ``ferromagnetic'' branch ($x\!\rightarrow\! +\infty$),  a
similar consideration yields the asymptotic scaling
$f_3(x)\!\sim\!x^{-0.33(1)}$,  indicated as a dashed line in figure
\ref{fig4}.  

Further, we found that the probability $P_L(\rho)$ that the MEDW
roughness is equal to $L$ scales as
$P_L(\rho)\!=\!f_4[(\rho-\rho_{\rm{c}})L^{1/\nu}]$, shown in the inset of  figure
\ref{fig4}.  In the ferromagnetic phase the value of $P_L$ tends towards zero
and in the spin-glass phase it saturates around $P_L\!\approx\!0.12$.  Hence,
as pointed out in \cite{krzakala2002}, an asymptotic nonzero probability that
the MEDW roughness is $O(L)$ can be used as an order parameter to  detect the
SG phase.
%

Regarding Model II, we simulated systems of size $L=24,32,48,64$ at different
values of the disorder parameter $r$. 
Here, we fixed the width of the disorder distribution to the 
value $\sigma_J\!=\!1$ and  we vary only its mean $\mu_J$.
Observables are averaged over up to
$3\!\times\!10^4$ ($2\!\times\!10^4$) samples for the smallest (largest)
systems and we utilized the data collapse anticipated by the scaling
assumptions for the Binder parameter  (see figure \ref{fig2ab}(b)) and the
magnetization (not shown) to obtain the values  $r_{\rm{c}}\!=\!0.970(2)$,
$\nu\!=\!1.49(4)$ ($S\!=\!1.0$) and $\beta\!=\!0.09(1)$ ($S\!=\!0.46$).  Note
that the numerical values of $r_{\rm c}$ and $\nu$ agree within errorbars with
the  values $r_{\rm c}\!=\!0.961(10)$ and $\nu\!=\!1.42(8)$ obtained using a
transfer-matrix approach \cite{mcmillan1984b}.  The scaling of the average
MEDW length here yields a  numerical value $d_{\rm f}\!=\!1.249(5)$
($S\!=\!1.99$) which can only be considered as an  effect of the finite system
size, see the discussion below. Further, the probability that the MEDW
roughness equals $L$ tends towards $P_L\!\approx\!0.12$ in the SG phase, in
agreement with the above results.
%
%

\subsection{Scaling behavior at fixed values of $\rho$ and $r$}
We have carried out further simulations at a couple of selected
  values of $\rho$ and $r$, see
tables \ref{tab1}  and \ref{tab2}, in order to probe the asymptotic scaling
behavior of MEDWs regarding the  two disorder distributions introduced above.
We therefore considered systems of size up to $L\!=\!512$ with $10^3$
realizations of the disorder.  In particular, we are interested in the
asymptotic scaling behavior of the average MEDW  length $\langle l \rangle$
with respect to the system size $L$, defining the  DW fractal dimension
$d_{\rm f}$ via $\langle l \rangle\!\sim\!L^{d_{\rm f}}$.  We further study
the scaling of the average MEDW roughness $\langle h \rangle$, i.e.\ the
extension of  the lattice path in the direction of the periodic BCs,  that
defines the roughness exponent $d_{\rm r}$ by means of $\langle
h\rangle\!\sim\! L^{d_{\rm r}}$.  Both these observables relate only to the
geometric properties of the MEDW, see figure \ref{fig1}.  Finally, we
investigate the size scaling of the mean $\Delta E \!=\! \langle |\delta E|
\rangle \!\sim\!L^{\theta_1}$ and  width $\sigma(\delta E)\!=\!\sqrt{\langle
\delta E^2 \rangle - \langle \delta E \rangle^2}\!\sim\!L^{\theta_2}$ of the
distribution of MEDW excitation  energies.

\begin{figure}[t!]
\centerline{
\includegraphics[width=1.0\linewidth]{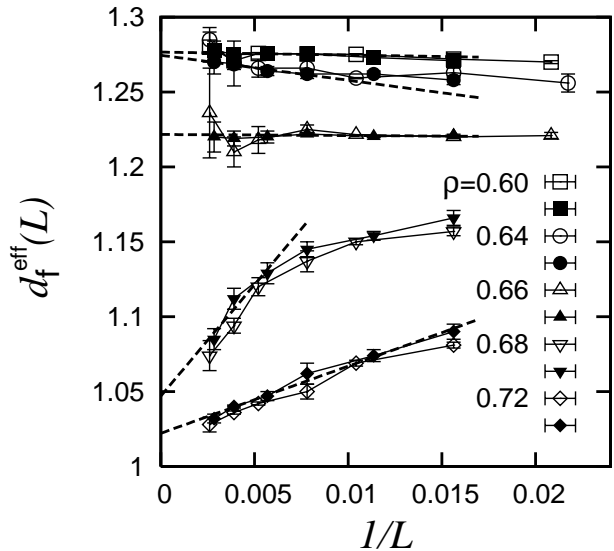}} 
\caption{Extrapolation of the asymptotic fractal dimension for Model I.
Analysis of the sequence of effective exponents $d_{\rm f}^{\rm eff}(L)$
that describe the scaling of the average MEDW length
within intervals of 3 (open symbols) and 4 (filled symbols) successive 
values of $L$, according to $\langle l \rangle\!\sim\!L^{d_{\rm f}}$.
The asymptotic value of the fractal dimension $d_{\rm f}$ is extrapolated
from the plot of $d_{\rm f}^{\rm eff}$ against $1/L$ as the intersection 
of a straight line fit to the data with the ordinate.
\label{fig5}}
\end{figure}

Again, we first discuss the results for the Model I disorder and afterwards state the results 
for the Model II disorder more briefly.
\begin{table}[b]
\begin{center}
\begin{tabular}[c]{l@{\quad}l@{\quad}l@{\quad}l@{\quad}l}
\hline
\hline
   $\rho$    & $d_{\rm f}$     & $d_{\rm r}$  & $\theta_1$& $\theta_2$  \\
\hline
0.00     & 1.274(2) & 1.008(11)& -0.287(4)& -0.287(4)\\
0.60     & 1.275(1) & 1.003(3) & -0.28(1) & -0.28(2) \\
0.64     & 1.275(2) & 1.012(4) & -0.28(1) & -0.28(4) \\
0.66     & 1.222(1) & 1.002(2) & 0.17(2)  & 0.16(1)  \\
0.68     & 1.05(2)  & 0.74(3)  & 0.97(4)  & 0.35(3)  \\
0.72     & 1.022(1) & 0.698(6) & 1.052(3) & 0.27(2)  \\
\hline
\hline
\end{tabular}
\end{center}
\caption{From left to right: 
disorder parameter, fractal dimension, 
roughness exponent and exponents that describe the 
scaling of the mean and width of the MEDW energy distribution.
The figures for $\rho\!=\!0$ are taken from \cite{melchert2007}.
\label{tab1}}
\end{table}
%
\begin{table}[b]
\begin{center}
\begin{tabular}[c]{l@{\quad}l@{\quad}l@{\quad}l@{\quad}l}
\hline
\hline
   $r$    & $d_{\rm f}$     & $d_{\rm r}$  & $\theta_1$& $\theta_2$  \\
\hline
$\infty$         & 1.274(2) & 1.008(11)& -0.287(4)& -0.287(4)\\
1.111    & 1.275(7) & 0.994(4)& -0.294(6)& -0.295(5)\\
1.010    & 1.286(3) & 1.024(2) & -0.311(2) & -0.35(1) \\
0.970    & 1.222(6) & 0.999(3) & 0.15(1) & 0.15(1) \\
0.935    & 1.085(4) & 0.782(3) & 0.96(2)  & 0.31(2)  \\
0.833    & 1.015(1)  & 0.651(3)  & 1.028(1)  & 0.31(2)  \\
\hline
\hline
\end{tabular}
\end{center}
\caption{From left to right: 
disorder parameter $r=\sigma_J/\mu_J$, fractal dimension, 
roughness exponent and exponents that describe the 
scaling of the mean and width of the MEDW energy distribution.
The figures for $r\!=\!\infty$, i.e.\ $\mu_J\!=\!0$ ,are taken from \cite{melchert2007}.
\label{tab2}}
\end{table}
%
The asymptotic scaling behavior of the average DW length
allows one to obtain the fractal dimension by using a direct fit to the power law data
over the entire range of system sizes $L$. A reliable and more systematic alternative
is to investigate a sequence of effective (local) exponents $d_{\rm f}^{\rm eff}(L)$ that
describe the scaling of $\langle l \rangle$ within intervals of, say, 3 successive
values of $L$. The change of the effective exponents for increasing system sizes further
show how the scaling behavior is affected by the finite size of the simulated systems.
From the sequence of effective exponents one can extrapolate the asymptotic fractal
dimension by means of a straight line fit to the plot of $d_{\rm f}^{\rm eff}(L)$ against
the inverse system size $1/L$. Figure \ref{fig5} shows the effective exponents obtained for
$3$ and $4$ successive values of $L$ at different values of the disorder parameter $\rho$. 
Therein, the asymptotic fractal dimensions $d_{\rm f}$, as listed in table \ref{tab1}, 
where estimated from the effective exponents resulting from intervals of 
$4$ successive system sizes.
The asymptotic values for $d_{\rm r}$, $\theta_1$ and $\theta_2$, listed in tables \ref{tab1}/\ref{tab2}, 
where estimated using a similar procedure.
Our results for the fractal dimension and the stiffness exponent at $\rho\!=\!0.60$ and $0.62$
clearly support the estimates for the pure SG at $\rho\!=\!0$.
They are in agreement with the SLE scaling relation and hence we could verify
that the SLE scaling relation holds up to values of the disorder parameter close
to $\rho_{\rm c}$.
%

At the critical point we find that the estimates of $d_{\rm f}$ and $\theta_2$ 
are not in agreement with the SLE scaling relation.
However, MEDWs at $\rho_{\rm c}$ are self-similar with the scaling
dimension $d_{\rm f}\!=\!1.222(1)$ and a roughness compatible with unity. Here,
the numerical value of $d_{\rm f}$ as estimated from the effective exponents 
compares nicely to the value $d_{\rm f}\!=\!1.222(4)$ found from the 
previous FSS.

We further find that $d_{\rm f}$ and $\theta_2$ in the ferromagnetic phase above 
the critical point are not consistent with the SLE scaling 
relation.
There, the overall length of the DW 
increases linear with the system size, i.e.\ 
the fractal dimension extrapolates towards $d_{\rm f}\!=\!1$, whereas 
for the roughness exponent $d_{\rm r}\!<\!1$ is found.
This indicates that, albeit the MEDW is allowed to 
bend and turn back and forth on the lattice, the resulting overhangs
are not significant for their scaling behavior. 
Further, the cost needed to introduce the DW grows almost linearly with 
the system size, while the rms--fluctuation is characterized by an 
exponent significantly smaller than that.
Hence, MEDWs in the ferromagnetic phase display a self-affine scaling, governed by
exponents that are in reasonable agreement with those
that describe the scaling of the transverse deviation ($\sim\!L^{2/3}$) and
the rms excitation energy ($\sim\!L^{1/3}$) of pinned DWs in an ordinary Ising FM with 
randomly placed impurities \cite{huse1985}. Further, the scaling 
behavior found here agrees with that observed for directed and 
undirected optimal paths on $2d$ lattices subject to weak disorder 
\cite{schwartz1998} or analogously the scaling of directed polymers in 
random media \cite{kardar1987}.

For the Model II disorder, our findings are qualitatively the same,
hence we only state the numerical results without showing figures.  The
numerical values of $d_{\rm f}$ and $\theta_2$ within the SG phase
($r\!>\!r_{\rm c}$) are in agreement with the SLE scaling relation proposed
for the pure SG. In particular, at $r\!=\!1.01$ we find $d_{\rm
f}\!=\!1.286(3)$ and $\theta_2\!=\!-0.35(1)$.
Here, the data for $\langle l \rangle$ gives a nice straight line on a 
double logarithmic scale, where we find $d_{\rm{f}}\!=\!1.284(2)$ from a fit to
the pure power law data excluding $L\leq100$. The situation for the
data corresponding to $\sigma(\delta E)$ is somewhat different, i.e. the 
data still exhibits a curvature within the range of accessible 
system sizes on a double logarithmic scale. This does not allow to 
fit all the data at once, assuming a power law fit-function. Consequently,
the most reliable estimate of the asymptotic value of $\theta_2$ can
be obtained by an analysis of the local exponents as described above.
The reason for this difficulty might stem from the fact that the value 
$r\!=\!1.01$ of the disorder parameter is located in the transition region 
close to the critical point.
Albeit these values differ slightly from the
values $\theta_2\!\approx\!-0.28$ and $d_{\rm f}\!\approx\!1.274$ that one
would expect  to find in the SG phase, they are in agreement with the SLE
scaling relation.  The numerical values for the exponents right at the
critical point are again not  in agreement with the proposed scaling
relation. However, the asymptotic  fractal dimension extrapolated from the
effective exponents reads $d_{\rm f}\!=\!1.222(6)$ and is in agreement with
the corresponding value at the critical point for the  Model I
disorder. Further, if we analyze the scaling of the average MEDW length
restricted to system sizes $L\!<\!64$ we find a value of $d_{\rm
f}\!=\!1.246(4)$, consistent with the value encountered in the previous FSS
analysis that was denoted as a finite-size effect.  Within the ferromagnetic
phase ($r\!<\!r_{\rm c}$), MEDWs again display a  self-affine scaling
behavior, further characterized by exponents that  extrapolate towards those
that describe  the scaling of the transverse deviation and the rms excitation
energy of pinned DWs in an ordinary Ising FM with  randomly placed impurities.
Referring to the droplet model and under the assumption that all lengths exhibit
the same asymptotic scaling behavior, one can further relate the exponents $d_{\rm{f}}$
and $\theta_2$ by means of the equation $d_{\rm{f}}=d/2-\theta_2$ \cite{liers2007}.
Note that within the SG phase our data is in reasonable agreement with this scaling 
relation. In this regard, the agreement for model I is somewhat better than for model II.
Further, combining the above equation with the SLE scaling relation \cite{fisch2007}, 
we would expect $\theta_2\!\approx\!-0.2753$ or similarly $d_{\rm{f}}\!\approx\!1.2753$.

%
\begin{figure}[t!]
\centerline{
\includegraphics[width=1.0\linewidth]{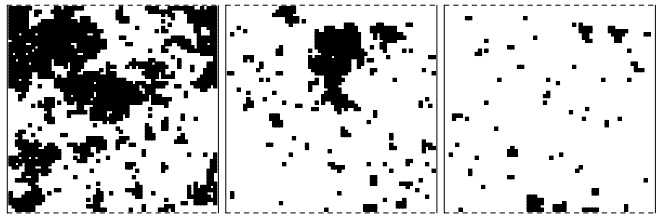}} 
\caption{
Samples of GS spin configurations for systems of side-length 
$L\!=\!64$ and $r\!=\!1.1,0.969,0.9$ (from left to right) for Model II.
\label{fig8}}
\end{figure}  
%
%
%
\begin{figure}[t!]
\centerline{
\includegraphics[width=1.0\linewidth]{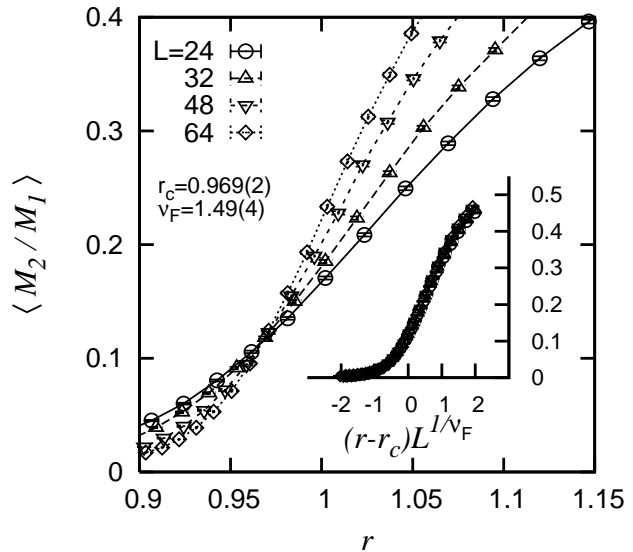}} 
\caption{FSS analysis of ferromagnetic domains at $T\!=\!0$ for Model II.
The main plot shows the average size-ratio $\langle M_2/M_1\rangle$ of the 
second-largest and largest ferromagnetic clusters and the inset illustrates 
the data collapse under the respective scaling assumption, obtained for 
$r_{\rm{c}}\!=\!0.969(2)$ and $\nu_{\rm{F}}\!=\!1.49(4)$.
\label{fig6}}
\end{figure}  
%
%
\begin{figure}[t!]
\centerline{
\includegraphics[width=1.0\linewidth]{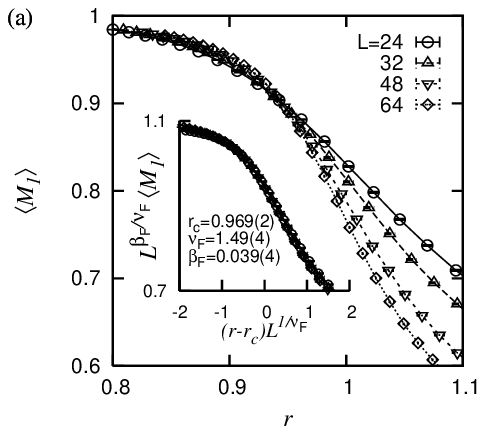}} 
\centerline{
\includegraphics[width=1.0\linewidth]{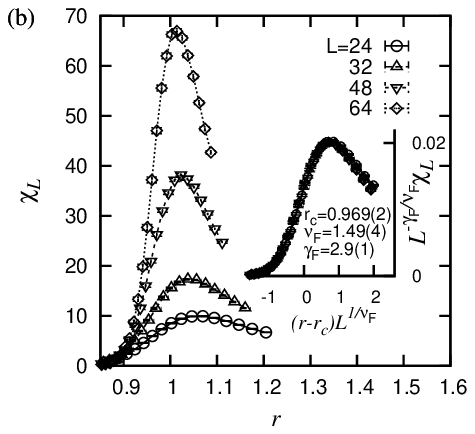}} 
\caption{FSS analysis of ferromagnetic domains at $T\!=\!0$ for Model II.
(a) normalized size $\langle M_1 \rangle$ of the largest ferromagnetic cluster, and
(b) finite-size susceptibility $\chi_L=N [\langle M_1^2 \rangle - \langle M_1 \rangle^2]$ associated with the 
the size of the largest cluster.
The main plots show the unscaled data near the critical
point, while the insets illustrate the data collapse under the 
respective scaling assumptions.
\label{fig7ab}}
\end{figure}  

\subsection{Finite-size scaling analysis of ferromagnetic spin domains at the $T\!=\!0$ spin 
glass to ferromagnet transition}

As we decrease the value of the disorder parameter in Model II from
$r\!=\!\infty$ (SG-phase)  to $r\!<\!r_{\rm{c}}$ (FM-phase), we can identify
ferromagnetic clusters of spins, i.e.\ groups of  nearest-neighbor
spins with similar orientation, with increasing size (see Fig.\
\ref{fig8}).  Here, as an alternative way to characterize the SG to FM
transition at $T\!=\!0$, we  perform a FSS analysis of the largest and
second largest ferromagnetic clusters found  for the GS spin
configuration for each realization of the disorder.  Such an
analysis has been performed previously for standard percolation
\cite{dasilva2002}. As above, we simulated systems of size
$L=24,32,48,64$ at different values of the disorder parameter $r$.  We
kept the width of the disorder distribution at the fixed value
$\sigma_J\!=\!1$ and  we vary only its mean $\mu_J$.  Observables are
averaged over $2\!\times\!10^4$ samples for each system size.
Subsequently, the relative size of a cluster specifies the number of
spins that comprise the cluster divided by the number of spins on the
lattice.  Within our analysis we found that the average ratio $\langle
M_2/M_1 \rangle$ of the relative sizes  of the second-largest and the
largest ferromagnetic clusters scales as
\begin{equation} 
\langle M_2/M_1 \rangle\!\sim\!f_5[(r-r_{\rm{c}})L^{1/\nu_{\rm{F}}}]\,,
\label{eq:scaling_F:1}
\end{equation}
therein $r_{\rm{c}}$ is the location of the critical point and $\nu_{\rm{F}}$
signifies the correlation  length exponent. From a data collapse,
restricted to the interval $[-1.5,+1.5]$ on the rescaled abscissa, we
obtain the numerical values $r_{\rm{c}}\!=\!0.969(2)$ and $\nu_F\!=\!1.49(4)$
with a quality $S\!=\!0.85$, see figure \ref{fig6}. Both values agree
within errorbars with those obtained from the Binder parameter
analysis. If we allow for a nonzero scaling dimension according to
$\langle M_2/M_1 \rangle\!\sim\! L^{-\kappa}
f_6[(r-r_{\rm{c}})L^{1/\nu_{\rm{F}}}]$, we yield  $r_{\rm{c}}$ and $\nu_F$ as above
and further $\kappa=0.004(13)$ ($S\!=\!0.82$, $[-2.0,+1.0]$).  The
numerical value of $\kappa$ is compatible with zero and hence supports
the scaling  assumption (\ref{eq:scaling_F:1}) for the size
ratio. Moreover, right at $r_{\rm{c}}$ we found the  critical value
$\langle M_2/M_1\rangle\!=\!0.122(1)$. For completeness we note, that
we yield qualitatively similar findings for the ratio $\langle
M_2\rangle /\langle M_1\rangle$ with  the critical value $\langle
M_2\rangle /\langle M_1\rangle\!=\!0.098(1)$.  The difference between
the two ratios is simply due to the cluster-size fluctuations at
criticality.

As an order parameter we measure the relative size $M_1$ of the
largest ferromagnetic cluster for each of the GSs. From the scaling
assumption  $\langle M_1
\rangle\!\sim\!L^{-\beta_{\rm{F}}/\nu_{\rm{F}}} f_7[(r-r_{\rm c})
L^{1/\nu_{\rm{F}}}]$ and the values of $r_{\rm{c}}$ and $\nu_{\rm{F}}$
stated above we obtain $\beta_{\rm{F}}\!=\!0.039(4)$ ($S\!=\!0.54$,
$[-0.5,+0.5]$), see figure \ref{fig7ab}(a).  A similar scaling
assumption for the second largest cluster yields
$\beta_{\rm{F},2}=0.05(3)$  ($S\!=\!0.26$, $[-0.5,+0.25]$, not shown).
Albeit the numerical value of $\beta_{\rm{F},2}$ is less precise and
somewhat larger compared to $\beta_{\rm{F}}$, both exponents are
compatible with Eq.\ (\ref{eq:scaling_F:1}).

The finite-size susceptibility $\chi_L\!=\!N [\langle M_1^2 \rangle
  -\langle M_1 \rangle^2]$ describing the fluctuations of the
size of the largest ferromagnetic cluster, obeys the scaling form
$\chi_L\!\sim\!L^{\gamma_{\rm{F}}/\nu_{\rm{F}}} f_8[(r-r_{\rm c})
  L^{1/\nu_{\rm{F}}}]$ with another critical exponent
$\gamma_{\rm{F}}$, see figure \ref{fig7ab}(b). Together with the
values of $r_{\rm{c}}$ and $\nu_{\rm{F}}$ we estimate
$\gamma_{\rm{F}}\!=\!2.9(1)$ ($S\!=\!0.85$, $[-1.5,+1.0]$).  These
exponents further are in agreement with the hyperscaling relation
$\gamma_{\rm{F}}/\nu_{\rm{F}} + 2 \beta_{\rm{F}}/\nu_{\rm{F}}=d$.

We performed further simulations for the $\pm$J model with a varying
fraction $0.0\leq p \leq 0.5$ of  aniferromagnetic bonds. In
principle, the GS for this model is highly degenerate
\cite{bray1986,landry2002}. Here, we  investigate only one randomly obtained GS
for each realization of the disorder.  From a FSS analysis for systems
of size $L\!=\!32,48,64,96$, where averages are computed  over
$3\!\times\! 10^4$ samples, we found $p_{\rm{c}}\!=\!0.1022(3)$,
$\nu_{\rm{F}}\!=\!1.47(6)$, $\beta_{\rm{F}}\!=\!0.037(4)$  and
$\gamma_{\rm{F}}\!=\!2.8(1)$. The numerical values of the critical
exponents for the $\pm$J model agree, within  errorbars, with those
obtained for Model II above. Further, the critical concentration of
antiferromagnetic bonds is in fair agreement with the value
$p_{\rm{c}}\!=\!0.103(1)$ found from an  analysis of the Binder
parameter within a previous study \cite{amoruso2004}.	 Regarding the
FSS analysis and compared to \cite{amoruso2004}, we used a larger
number of  interpolation points that enclose the critical point on the
rescaled abscissa  ($24$ data points in the interval $[-0.5:0.5]$ for
each system size).  As a result we obtained $p_{\rm{c}}$ with
increased precision, although our  system sizes are somewhat smaller.
Finally, right at $p_{\rm{c}}$ we found the critical ratios $\langle
M_2/M_1 \rangle\!=\!0.104(1)$  and $\langle M_2\rangle /\langle
M_1\rangle\!=\!0.083(1)$. The numerical values of these ratios differ
slightly from those obtained for Model II above. However, in both
cases  we observe $\langle M_2/M_1\rangle \!\approx \!0.125\, \langle
M_2\rangle /\langle M_1\rangle$.

As mentioned above, the scaling of the size ratio according
to equation \ref{eq:scaling_F:1} was also confirmed for usual random
percolation \cite{dasilva2002}. It stems from the fact that  the
largest and second-largest clusters exhibit the same fractal dimension
at the critical point.  For usual percolation this was shown earlier
\cite{jan1998}.  While we could verify equation (\ref{eq:scaling_F:1})
for the disorder induced SG to  FM transition at $T\!=\!0$
numerically, we found within additional simulations no such
scaling behavior for the thermal phase transition in the 2d  Ising
ferromagnet.


\section{Summary}
\label{sec:summary}
We have investigated MEDWs for two-dimensional random-bond Ising spin
systems,  regarding two different continuous bond distributions. For
both models, a disorder parameter could be used to distinguish between
a spin-glass ordered or a  ferromagnetic ground state. We performed a
FSS analysis to locate the  critical points in both models that
separate the spin-glass phase from  the ferromagnetic phase. We found
that within the spin-glass phase, the  exponents that describe the
size scaling of the width of the average DW  energy and the average DW
length are approximately constant  and consistent with the SLE scaling
relation previously proposed for the  pure spin-glass.  Right at the
critical point and in the ferromagnetic phase of the models the
accordant exponents are not in agreement with the SLE scaling relation.

It is intriguing to note that the fractal dimension of the DWs at the
critical point of both disorder types studied here, agrees with the
fractal dimension $d_{\rm opt}=1.22(2)$ of optimal paths in the strong
disorder limit on $2d$ lattices \cite{cieplak1994}.  This is quite
interesting since the optimization criteria of the two problems are
rather distinct: In the strong disorder limit, nonnegative edge weights
are drawn from a very broad distribution. The cost of a path between
two sites on the lattice is  then dominated by the largest edge-weight
along the path. Consequently,  so as to find an optimal path, one has
to minimize the largest weight along the path.  In contrast to this,
the cost of a MEDW is the sum of all edge weights along the respective
lattice path. There are positive and also negative edge weights that
can cancel each other, at least partially.  A common feature of the
above two problems is that, in striking contrast to usual shortest
path problems, there is no immediate negative feedback for the
inclusion of additional path segments. In usual shortest path
problems, where there are only positive edge weights, like e.g.\
optimal paths subject to  weak disorder \cite{schwartz1998}, the
inclusion of additional path segments leads very  likely to  an
increased path weight. Hence, postive-weight minimum-weight paths tend
to be short, which results in an average end-to-end distance $\sim\!L$.

Finally, we have characterized the SG to FM transition at $T\!=\!0$ in
terms of the  largest and second-largest ferromagnetic clusters of
spins found for the GS spin configurations. The respective critical
exponents support our previous  results and they appear to be
consistent with a hyperscaling relation known from  scaling theory.\\[0.5cm]
After submission of our manuscript to Phys. Rev. B, we received 
correspondence from R. Fisch and E. Vicari containing valuable 
suggestions and comments that enabled us to improve the manuscript 
further. 
The article was accepted without a request for amendments. Hence
and to our regret, these improvements do not appear in the published 
version [Phys. Rev. B {\bf 79}, 184402 (2009)] of the article.

\begin{acknowledgments}
We acknowledge financial support from the VolkswagenStiftung (Germany)
within the program ``Nachwuchsgruppen an Universit\"aten''. 
 The simulations were performed at the workstation cluster of the ``Institute for
Theoretical Physics'' in G\"ottingen (Germany) and the GOLEM I cluster for scientific 
computing at the University of Oldenburg (Germany).\\[0.3cm]
In addition, we thank R. Fisch and E. Vicari for valuable suggestions and 
comments that enabled us to improve the manuscript further.
\end{acknowledgments}         

\bibliography{lit_DWscaling_2dSGFM.bib}

\end{document}